%% file: main.tex
\documentclass[runningheads]{llncs}
\usepackage[utf8]{inputenc}
\usepackage{graphicx}
\usepackage{bm}
\usepackage{xcolor}
\usepackage{xparse}
\usepackage{tikz}
\usetikzlibrary{shapes,arrows,positioning}
\usetikzlibrary{calc}
\usepackage{amsmath}
\usepackage{amsfonts}
\usepackage[binary-units=true]{siunitx}
\usepackage{booktabs, multirow}
\usepackage{mathtools}
\usepackage{hyperref}
\usepackage[capitalise]{cleveref}
\usepackage{nomencl}
\usepackage{multicol}
\usepackage{etoolbox}
\usepackage{adjustbox}
\usepackage{mathrsfs}
\usepackage{float}
\usepackage[shortlabels]{enumitem}
\usepackage[nofillcomment,linesnumbered,vlined,boxed,commentsnumbered]{algorithm2e}

\renewcommand{\nompreamble}{The next list describes important symbols used within the body of the document.}

\newsavebox{\tempbox}

\input{macros}

\begin{document}
\title{Power of Pre-Processing: Production Scheduling with Variable Energy Pricing and Power-Saving States}
\titlerunning{Scheduling with Energy Costs and Machine States}
\author{Ondřej Benedikt\inst{1,2} \and István Módos \inst{1,2} \and Zdeněk Hanzálek \inst{1}
}
\authorrunning{Benedikt et al.}
\institute{Czech Institute of Informatics, Robotics and Cybernetics, Czech Technical University in Prague, Czech Republic \and
Faculty of Electrical Engineering, Czech Technical University in Prague, Czech Republic \\
\email{\{ondrej.benedikt,istvan.modos,zdenek.hanzalek\}@cvut.cz}}
\maketitle
\begin{abstract} \textit{This is a pre-print of an article published in Constraints. The final authenticated version is available online at: \href{https://doi.org/10.1007/s10601-020-09317-y}{https://doi.org/10.1007/s10601-020-09317-y}}.

This paper addresses a single machine scheduling problem with non-preemptive jobs to minimize the total electricity cost.
Two latest trends in the area of the energy-aware scheduling are considered, namely the variable energy pricing and the power-saving states of a machine.
Scheduling of the jobs and the machine states are considered jointly to achieve the highest possible savings.
Although this problem has been previously addressed in the literature, the reported results of the state-of-the-art method show that the optimal solutions can be found only for instances with up to 35 jobs and 209 intervals within 3 hours of computation.
We propose an elegant pre-processing technique called \AbbrIwa{} for computing the optimal switching of the machine states with respect to the energy costs.
The optimal switchings are associated with the shortest paths in an interval-state graph that describes all possible transitions between the machine states in time. 
This idea allows us to implement efficient integer linear programming and constraint programming models of the problem while preserving the optimality.
The efficiency of the models lies in the simplification of the optimal switching representation.
The results of the experiments show that our approach outperforms the existing state-of-the-art exact method.
On a set of benchmark instances with varying sizes and different state transition graphs, the proposed approach finds the optimal solutions even for the large instances with up to 190 jobs and 1277 intervals within an hour of computation. 

\end{abstract}
\section{Introduction}

Energy-efficient scheduling has been attracting a considerable amount of attention lately, as reported in both \cite{2016:Gahm} and \cite{2019:Gao}.
The trend is most likely to continue in the future since the energy-efficient scheduling helps to achieve sustainability of the production by both decreasing the production cost and minimizing its environmental impact.
Gahm et al. in \cite{2016:Gahm} identified promising approaches to the energy-aware scheduling, including, among others, (i) the optimization of the energy demand by considering the power-saving states of the machines, and (ii) the participation in demand response programs, which are used by the electric utilities to reward the energy consumers for shifting their energy consumption to \emph{off-peak} intervals \cite{2015:Merkert}.

In this work, we study a single machine scheduling problem to minimize the total energy cost (TEC) of the production.
We consider both the power-saving states of the machine and the time-of-use (TOU) pricing.
The TOU pricing is one of the demand response programs, in which the electricity price may differ every hour. 
The scheduling problems with TOU pricing have been extensively addressed in the literature \cite{2016:Fang,2019:Gong,2015:Hadera}.

Considering the power-saving states of the machine, Mouzon et al. in \cite{2007:mouzon} identified  that a significant energy cost reduction can be attained. 
However, the switchings between the machine states need to be planned carefully because of their non-negligible energy costs and transition times. 

The integration of the power-saving states and the TOU pricing was originally proposed by Shrouf et al. \cite{2014:Shrouf}, who designed an integer linear programming (\AbbrILP{}) model for the single machine problem with the fixed order of the jobs. However, it was proven in \cite{2019:Aghelinejad} that the problem with the fixed order of the jobs can be solved in polynomial time. Aghelinejad et al. \cite{2017:Aghelinejad} improved and generalized the existing \AbbrILP{} model to consider even an arbitrary order of the jobs, in which case the problem is $\mathcal{NP}$-hard \cite{2019:Aghelinejad}. However, 
in both \cite{2017:Aghelinejad} and \cite{2014:Shrouf}, only small instances of the problem have been solved optimally. 

In this paper, we describe a novel pre-processing technique for the single machine scheduling problem, which was introduced in \cite{2014:Shrouf} and further studied in \cite{2017:Aghelinejad}. 
Our pre-processing technique pre-computes the optimal switching behavior in time w.r.t. energy costs. 
The pre-computed costs of the optimal switchings allow us to design efficient exact \AbbrILP{} and constraint programming (\AbbrCP{}) models.
In contrast, the \AbbrILP{} model proposed in~\cite{2017:Aghelinejad} explicitly formulates the transition behavior of the machine, which needs to be optimized jointly with the scheduling of the jobs.
As shown by the experiments, our approach outperforms the existing \AbbrILP{} model~\cite{2017:Aghelinejad}, which is, to the best of our knowledge, the state-of-the-art among the exact methods for this problem.
Our \AbbrILP{} model is able to solve all the benchmark instances with up to 190 jobs and 1277 pricing intervals within the time-limit.
On the other hand, the state-of-the-art \AbbrILP{} model from the literature scales only up to instances with 60 jobs and 316 intervals.

\section{Problem Statement} \label{sec:problem-statement}

Let $\SetIntervals = \{\Interval{1}, \Interval{2}, \dots, \Interval{\NumIntervals}\}$ be a set of \DefTerm{intervals}, which partition the scheduling horizon.
The \DefTerm{energy costs} for the intervals are given by the vector $\EnergyCostVector = (\EnergyCost{1}, \EnergyCost{2}, \dots, \EnergyCost{\NumIntervals})$, where $\EnergyCost{\IdxInterval} \in \SetIntNonNeg$ is the energy (electricity) cost associated with interval $\Interval{\IdxInterval}$.
It is assumed, that every interval is one time unit long, i.e., $\Interval{1} = [0,1)$, $\Interval{2} = [1,2)$, $\dots$, $\Interval{\NumIntervals} = [\NumIntervals-1, \NumIntervals)$. 
Note that the physical representation of the time unit length can be different depending on the required granularity of the scheduling horizon. 

Let $\SetJobs = \{\Job{1}, \Job{2}, \dots, \Job{\NumJobs} \}$ be a set of \DefTerm{jobs}, which must be scheduled on a single machine, that is available throughout the whole scheduling horizon; we assume that \( \NumJobs \ge 1 \).
Each job $\Job{\IdxJob}$ is characterized by its \DefTerm{processing time} $\ProcTime{\IdxJob} \in \SetIntPos$, given in the number of intervals.
Scheduling of the jobs is non-preemptive, and the machine can process at most one job at the time.
All the jobs are available at the beginning of the scheduling horizon.

During each interval, the machine is operating in one of its \DefTerm{states} $\IdxState \in \SetStates$ or transits from one state to another.
Let us denote the \DefTerm{transition time function} by $\TransTime : \SetStates \times \SetStates \rightarrow \SetIntNonNeg \cup \{\infty\}$, and the \DefTerm{transition power function} by \hbox{$\TransPower : \SetStates \times \SetStates \rightarrow \SetIntNonNeg \cup \{ \infty \}$}. 
The transition from state $\IdxState$ to state $\IdxAnother{\IdxState}$ lasts $\TransTime[\IdxState,\IdxAnother{\IdxState}]$ intervals and has power consumption  $\TransPower[\IdxState,\IdxAnother{\IdxState}]$, which is the constant rate of the consumed energy at every time unit.
The value $\infty$ means that the direct transition does not exist.
We assume that $\TransPower[\IdxState,\IdxState]$ denotes the power consumption of the machine while staying in state $\IdxState$ for the duration of one interval.

Note that the transition time/power functions are general enough to represent many kinds of machines, e.g., those studied in \cite{2017:Aghelinejad,2017:Benedikt,2007:mouzon,2014:Shrouf}.

During the first and the last interval, the machine is assumed to be in off state $\StateOff \in \SetStates$. Besides, the machine has a single processing state, $\StateProc \in \SetStates$, which must be active during the processing of the jobs. 
Due to the transition from/to the initial/last $\StateOff$ state, the machine cannot be in $\StateProc$ state during the early/late intervals. Hence, we denote the \DefTerm{earliest} and the \DefTerm{latest} interval during which the machine can be in $\StateProc$ state by $\Interval{\FirstOn}$ and $\Interval{\LastOn}$, respectively.

A \DefTerm{solution} is a pair $(\SolVecStarts, \SolVecTrans)$, where $\SolVecStarts = (\SolStarts{1}, \SolStarts{2}, \dots, \SolStarts{\NumJobs}) \in \SetIntNonNeg^{\NumJobs}$ is the vector denoting the start time of the jobs, and $\SolVecTrans = (\SolTrans{1}, \SolTrans{2}, \dots, \SolTrans{\NumIntervals}) \in (\SetStates \times \SetStates)^{\NumIntervals}$ represents the active state or transition in each interval.
The solution is feasible if the following four conditions are satisfied.

\begin{enumerate}
    \item the machine processes at most one job at a time;
    \item the jobs are processed when the machine is in $\StateProc$ state, i.e., \\ \hbox{$\forall \Job{\IdxJob} \in \SetJobs \  \forall \IdxInterval \in \IntInterval{\SolStarts{\IdxJob}+1}{\SolStarts{\IdxJob} + \ProcTime{\IdxJob}}: \SolTrans{\IdxInterval} = (\StateProc, \StateProc)$,}
    
    where $\IntInterval{a}{b}$ is $\{a, a+1, \dots, b\}$;
    \item the machine is in $\StateOff$ state during the first and the last interval, i.e., \\
    $\SolTrans{1} = (\StateOff, \StateOff)$, and $\SolTrans{\NumIntervals} = (\StateOff, \StateOff)$;
    
    \item all transitions are valid with respect to the transition time function. \\
\end{enumerate}
The total energy cost (TEC) of solution $(\SolVecStarts, \SolVecTrans)$ is

\begin{equation} \label{eq:tec}
    \sum\limits_{\Interval{\IdxInterval} \in \SetIntervals} \EnergyCost{\IdxInterval} \cdot \TransPower[\SolTrans{\IdxInterval}],
\end{equation}
where $\TransPower[\SolTrans{\IdxInterval}]$ represents $\TransPower[\IdxState,\IdxAnother{\IdxState}]$ for $\SolTrans{\IdxInterval} = (\IdxState, \IdxAnother{\IdxState})$.
The goal of the scheduling problem is to find a feasible solution minimizing the total energy cost \eqref{eq:tec}.

The above-defined problem was introduced in \cite{2014:Shrouf} and is denoted in standard Graham's notation as $\ProblemNotation$. 
The problem was shown to be strongly $\mathcal{NP}$-hard, see \cite{2019:Aghelinejad}.

\begin{figure}[h!]
    \centering
    \begin{tabular}{c@{\hskip 0.5cm}c@{\hskip 0.5cm}c}
	\includegraphics{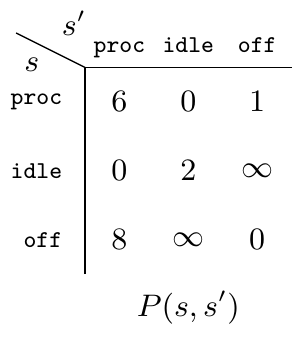} &
	\includegraphics{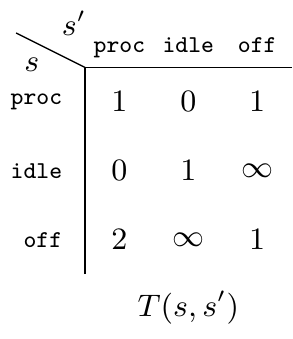} &	
	\includegraphics{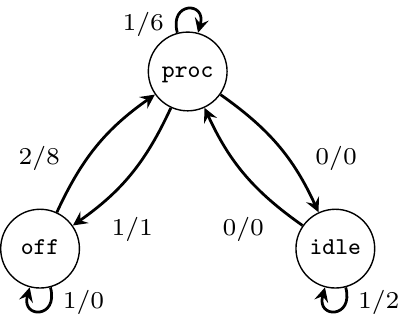} \\
    \end{tabular}
    \caption{Parameters of the transition power function \( \TransPower[\IdxState, \IdxAnother{\IdxState}] \) and transition time function \( \TransTime[\IdxState, \IdxAnother{\IdxState}] \), and the corresponding transition graph, where every edge from $\IdxState$ to $\IdxAnother{\IdxState}$ is labeled by $\TransTime[\IdxState, \IdxAnother{\IdxState}]$/$\TransPower[\IdxState, \IdxAnother{\IdxState}]$.}
    \label{fig:example-func-power-time}
\end{figure}

\paragraph{Example:} Here, we present a small example to illustrate the proposed notation.
Let us consider a scheduling horizon consisting of 16 intervals, \hbox{$\SetIntervals = \{\Interval{1}, \dots, \Interval{16}\}$}, and the associated energy costs $\EnergyCostVector = (2,1,2,1,8,16,14,3,2,5,3,10,3,2,1,2)$.
Let us have three jobs, $\SetJobs = \{\Job{1}, \Job{2}, \Job{3} \}$ with processing times $\ProcTime{1} = 2$, $\ProcTime{2} = 1$, and $\ProcTime{3} = 2$. 
Considering the machine states, we assume $\SetStates = \{\StateProc, \StateOff, \StateIdle\}$.
The values of the transition time function and the transition power function are given in \cref{fig:example-func-power-time}.
For the given transition time function, we have $\Interval{\FirstOn} = \Interval{4}$ and $\Interval{\LastOn} = \Interval{14}$.
Note that the same  machine states and transitions were originally proposed in \cite{2014:Shrouf}.

\begin{figure}[h]
    \centering
	\includegraphics{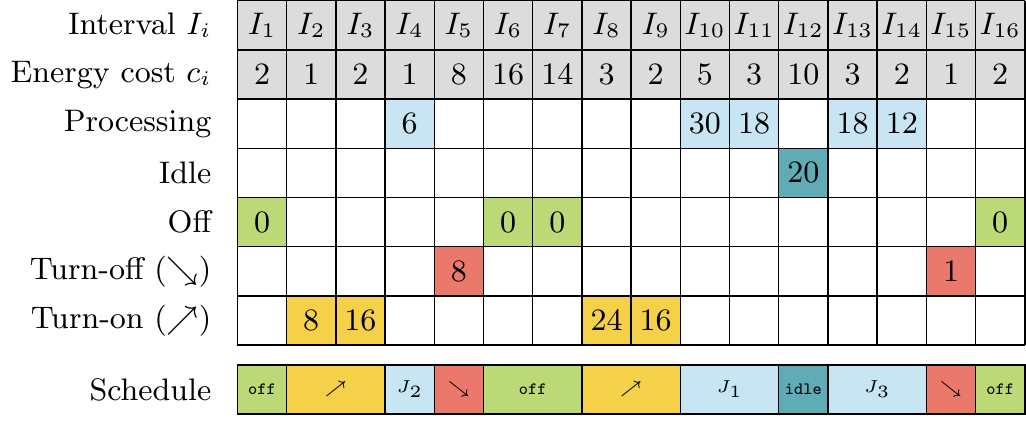}
    \caption{
    The optimal schedule for the example instance.
    Each cell, corresponding to interval $\Interval{\IdxInterval}$ and a state/transition, contains the value \( \EnergyCost{\IdxInterval} \cdot \TransPower[\SolTrans{\IdxInterval}] \).
    The sum over all these values gives the TEC equal to 177.
    }
    \label{fig:example-schedule}
\end{figure}

\noindent The optimal solution to the given instance is depicted in \cref{fig:example-schedule}, where
\begin{small}
\begin{align*}
\SolVecStarts & = (9, 3, 12), \ \text{and} \\
\SolVecTrans & =  ((\StateOff, \StateOff),
                  (\StateOff, \StateProc),
                  (\StateOff, \StateProc),
                  (\StateProc, \StateProc), 
                  (\StateProc, \StateOff),
                  (\StateOff, \StateOff), \\
                & \phantom{{}={}} (\StateOff, \StateOff),
                  (\StateOff, \StateProc),
                  (\StateOff, \StateProc),
                   (\StateProc, \StateProc), 
                 (\StateProc, \StateProc),
                  (\StateIdle, \StateIdle),  \\
                & \phantom{{}={}} (\StateProc, \StateProc),
                  (\StateProc, \StateProc),
                  (\StateProc, \StateOff),
                  (\StateOff, \StateOff)).
\end{align*}
\end{small}
The TEC of the optimal solution is equal to 177.

\section{Solution Approach} \label{sec:solution-approach}

In this section, we first describe how to pre-compute the optimal switching behavior of the machine and the corresponding costs. Afterward, we design efficient \AbbrILP{} and \AbbrCP{} models (called \AbbrIlpOur{} and \AbbrCpOur{}) that integrate the pre-computed optimal switching costs.

\subsection{Instance Pre-processing: Computation of the Optimal Switching}
\label{sec:pre-processing}

Given two states \( \IdxState, \IdxAnother{\IdxState} \) in which the machine is during two intervals \( \Interval{\IdxInterval}, \Interval{\IdxAnother{\IdxInterval}} \) such that \( \IdxInterval <  \IdxAnother{\IdxInterval} \), the pre-processing computes the optimal transitions from \( (\IdxState, \Interval{\IdxInterval}) \) to \( (\IdxAnother{\IdxState}, \Interval{\IdxAnother{\IdxInterval}}) \) over all possible states w.r.t.\ the energy cost.
Formally, the pre-processing solves the following optimization problem
\begin{equation} \label{eq:optimal-switching-problem}
    \min_{\SolTrans{\IdxInterval + 1}, \SolTrans{\IdxInterval + 2}, \dots, \SolTrans{\IdxAnother{\IdxInterval} - 1}} \sum_{\IdxAnotherTwo{\IdxInterval} = \IdxInterval + 1}^{\IdxAnother{\IdxInterval} - 1} \EnergyCost{\IdxAnotherTwo{\IdxInterval}} \cdot \TransPower[\SolTrans{\IdxAnotherTwo{\IdxInterval}}].
\end{equation}
such that \( ((\IdxState, \IdxState), \SolTrans{\IdxInterval + 1}, \SolTrans{\IdxInterval + 2}, \dots, \SolTrans{\IdxAnother{\IdxInterval} - 1}, (\IdxAnother{\IdxState}, \IdxAnother{\IdxState}) ) \) are valid transitions w.r.t.\ to the transition time function.
We call this an \DefTerm{optimal switching problem}.
As an illustration, the cost of the optimal switching in \cref{fig:example-schedule} from $ (\StateProc, \Interval{4}) $ to $ (\StateProc, \Interval{10}) $ equals 48.
Interestingly, the optimal switching problem can be solved in polynomial time by finding the shortest path in an \DefTerm{interval-state} graph, which is explained in the rest of this section.

The interval-state graph is defined by a triplet \( (\SetVertTrans, \SetEdgesTrans, \WeightTrans) \), where \( \SetVertTrans \) is the set of \DefTerm{vertices}, \( \SetEdgesTrans \) is the set of \DefTerm{edges} and \( \WeightTrans: \SetEdgesTrans \rightarrow \SetIntNonNeg \) are the \DefTerm{weights} of the edges.
The set of the vertices and edges of this graph are defined as follows:
\begin{align}
    \SetVertTrans & = \{ \VertTrans{1}{\StateOff} \} \cup \{ \VertTrans{\IdxInterval}{\IdxState} : \Interval{\IdxInterval} \in \SetIntervals \setminus \{ \Interval{1} \}, \IdxState \in \SetStates \} \cup \{ \VertTrans{\NumIntervals + 1}{\StateOff} \}, \\
    \begin{split}
        \SetEdgesTrans & = \{ (\VertTrans{1}{\StateOff}, \VertTrans{2}{\StateOff}) \}  \\
        & \phantom{{}={}} \cup \{ (\VertTrans{\IdxInterval}{\IdxState}, \VertTrans{\IdxInterval + \TransTime[\IdxState,\IdxAnother{\IdxState}]}{\IdxAnother{\IdxState}}) : \IdxState,\IdxAnother{\IdxState} \in \SetStates,  \Interval{\IdxInterval} \in \SetIntervals \setminus \{ \Interval{1} \}, \\
        & \phantom{{}={} \cup \{ (\VertTrans{\IdxInterval}{\IdxState}, \VertTrans{\IdxInterval + \TransTime[\IdxState,\IdxAnother{\IdxState}]}{\IdxAnother{\IdxState}}) : \,} \TransTime[\IdxState,\IdxAnother{\IdxState}] \not= \infty, (\IdxInterval - 1) + \TransTime[\IdxState,\IdxAnother{\IdxState}]  \le \NumIntervals - 1 \} \\
        & \phantom{{}={}} \cup \{  (\VertTrans{\NumIntervals}{\StateOff}, \VertTrans{\NumIntervals + 1}{\StateOff}) \}\,.
    \end{split}
\end{align}
Informally, each vertex \( \VertTrans{\IdxInterval}{\IdxState} \in \SetVertTrans \) represents that at the beginning of interval \( \Interval{\IdxInterval} \) the machine is in state \( \IdxState \).
Each edge \( (\VertTrans{\IdxInterval}{\IdxState}, \VertTrans{\IdxAnother{\IdxInterval}}{\IdxAnother{\IdxState}}) \in \SetEdgesTrans \) corresponds to the direct transition from state \( \IdxState \) to state \( \IdxAnother{\IdxState} \) that lasts \( \TransTime[\IdxState,\IdxAnother{\IdxState}] = (\IdxAnother{\IdxInterval} - \IdxInterval) \) intervals.
The condition \( (\IdxInterval - 1) + \TransTime[\IdxState,\IdxAnother{\IdxState}]  \le \NumIntervals - 1 \) ensures, that only transitions completing at most at the beginning of interval \( \Interval{\NumIntervals} \) are present in the interval-state graph.

The edges are weighted by the total energy cost of the corresponding transition w.r.t. the costs of energy in intervals, i.e., 
\DefTerm{weight} of edge \( (\VertTrans{\IdxInterval}{\IdxState}, \VertTrans{\IdxAnother{\IdxInterval}}{\IdxAnother{\IdxState}}) \in \SetEdgesTrans \) is defined as
\begin{equation}
    \WeightTrans[\VertTrans{\IdxInterval}{\IdxState}, \VertTrans{\IdxAnother{\IdxInterval}}{\IdxAnother{\IdxState}}] = \sum_{\IdxAnotherTwo{\IdxInterval} = \IdxInterval}^{\IdxAnother{\IdxInterval} - 1} \EnergyCost{\IdxAnotherTwo{\IdxInterval}} \cdot \TransPower[\IdxState, \IdxAnother{\IdxState}] \,.
\end{equation}

Returning to the optimal switching problem \eqref{eq:optimal-switching-problem}, the optimal transitions from \( (\IdxState, \Interval{\IdxInterval}) \) to \( (\IdxAnother{\IdxState}, \Interval{\IdxAnother{\IdxInterval}}) \) w.r.t.\ the energy cost can be obtained by finding the shortest path from \( \VertTrans{\IdxInterval + 1}{\IdxState} \) to \( \VertTrans{\IdxAnother{\IdxInterval}}{\IdxAnother{\IdxState}} \) in the interval-state graph.
We denote the cost of the optimal switching by function \( \PathCostTrans: \SetVertTrans \times \SetVertTrans \rightarrow \SetIntNonNeg \).
The values of \( \PathCostTrans \) can be computed using the Floyd-Warshall algorithm in \( \AsymCompUpper{\NumIntervals^3 \cdot \NumStates^3} \) time.

However, for the scheduling decisions, only some of the switchings are interesting.
Since all the jobs need to be scheduled in the $\StateProc$ state of the machine, the optimal switchings need to be resolved only in the `space', i.e., the sequence of intervals: (i) between two consecutive intervals with \( \StateProc \); (ii) between the first $\StateOff$ and the first \( \StateProc \); and (iii) the last \( \StateProc \) and the last $\StateOff$.
The cost of the switchings between \( \IdxState, \IdxAnother{\IdxState} \in \{ \StateOff, \StateProc\}^2\) are recorded by function \( \OptCostTrans : \SetIntervals^2 \rightarrow \SetIntNonNeg \) defined as
\begin{equation}
    \OptCostTrans[\IdxInterval, \IdxAnother{\IdxInterval}]=
    \left\lbrace \begin{aligned}
         & \PathCostTrans[\VertTrans{\IdxInterval + 1}{\StateProc}, \VertTrans{\IdxAnother{\IdxInterval}}{\StateProc}] & \IdxInterval > 1, \IdxAnother{\IdxInterval} < \NumIntervals & \quad\quad \text{\footnotesize{case (i)}} \\
         & \PathCostTrans[\VertTrans{2}{\StateOff}, \VertTrans{\IdxAnother{\IdxInterval}}{\StateProc}] & \IdxInterval = 1, \IdxAnother{\IdxInterval} < \NumIntervals & \quad\quad \text{\footnotesize{case (ii)}} \\
         & \PathCostTrans[\VertTrans{\IdxInterval + 1}{\StateProc}, \VertTrans{\NumIntervals}{\StateOff}] & \IdxInterval > 1, \IdxAnother{\IdxInterval} = \NumIntervals & \quad\quad \text{\footnotesize{case (iii)}}
    \end{aligned} \right.
\end{equation}
for each \( \IdxInterval < \IdxAnother{\IdxInterval} \). The vector of states corresponding to $\OptCostTrans[\IdxInterval, \IdxAnother{\IdxInterval}]$, i.e., the \DefTerm{optimal switching behavior} of the machine between $\IdxInterval$ and $\IdxAnother{\IdxInterval}$, is denoted by $\OptPathTrans[\IdxInterval, \IdxAnother{\IdxInterval}]$.
As an example, see the \cref{fig:example-schedule}, where intervals \( \{ \Interval{5}, \Interval{6}, \dots, \Interval{9} \}\) represent the space between two consecutive jobs \( \Job{2}, \Job{1} \) with cost \( \OptCostTrans[4,10] = 48 \).

Values of \( \OptCostTrans \) can be computed efficiently using an algorithm that we call the \DefTerm{Shortest Path Algorithm for Cost Efficient Switchings} (\AbbrIwa{}).
In every iteration \( \Interval{\IdxInterval} \in \SetIntervals \setminus \{ \Interval{\NumIntervals} \} \), \AbbrIwa{} computes all values \( \OptCostTrans[\IdxInterval, \IdxInterval + 1], \OptCostTrans[\IdxInterval, \IdxInterval + 2], \dots, \OptCostTrans[\IdxInterval, \NumIntervals] \) by finding the shortest paths from \( \VertTrans{\IdxInterval + 1}{\StateProc} \) (or \( \VertTrans{2}{\StateOff} \) if \( \IdxInterval = 1 \)) to all other vertices in the interval-state graph.
The shortest paths are obtained with Dijkstra algorithm that runs in \( \AsymCompUpper{|\SetEdgesTrans| + |\SetVertTrans| \cdot \log |\SetVertTrans|}\) if implemented using the priority queues.
Since the Dijkstra algorithm is started \( \NumIntervals \) times, the complexity of \AbbrIwa{} is
\begin{equation}
    \AsymCompUpper{\NumIntervals \cdot ( |\SetEdgesTrans| + |\SetVertTrans| \cdot \log |\SetVertTrans|)} = \AsymCompUpper{\NumIntervals^2 \cdot \NumStates \cdot (\NumStates + \log \NumIntervals \cdot \NumStates)}\,.
\end{equation}
Moreover, to increase the performance further, iterations \( \IdxInterval \) can be computed in parallel since they are independent to each other.

\begin{figure}[t]
    \centering
	\includegraphics{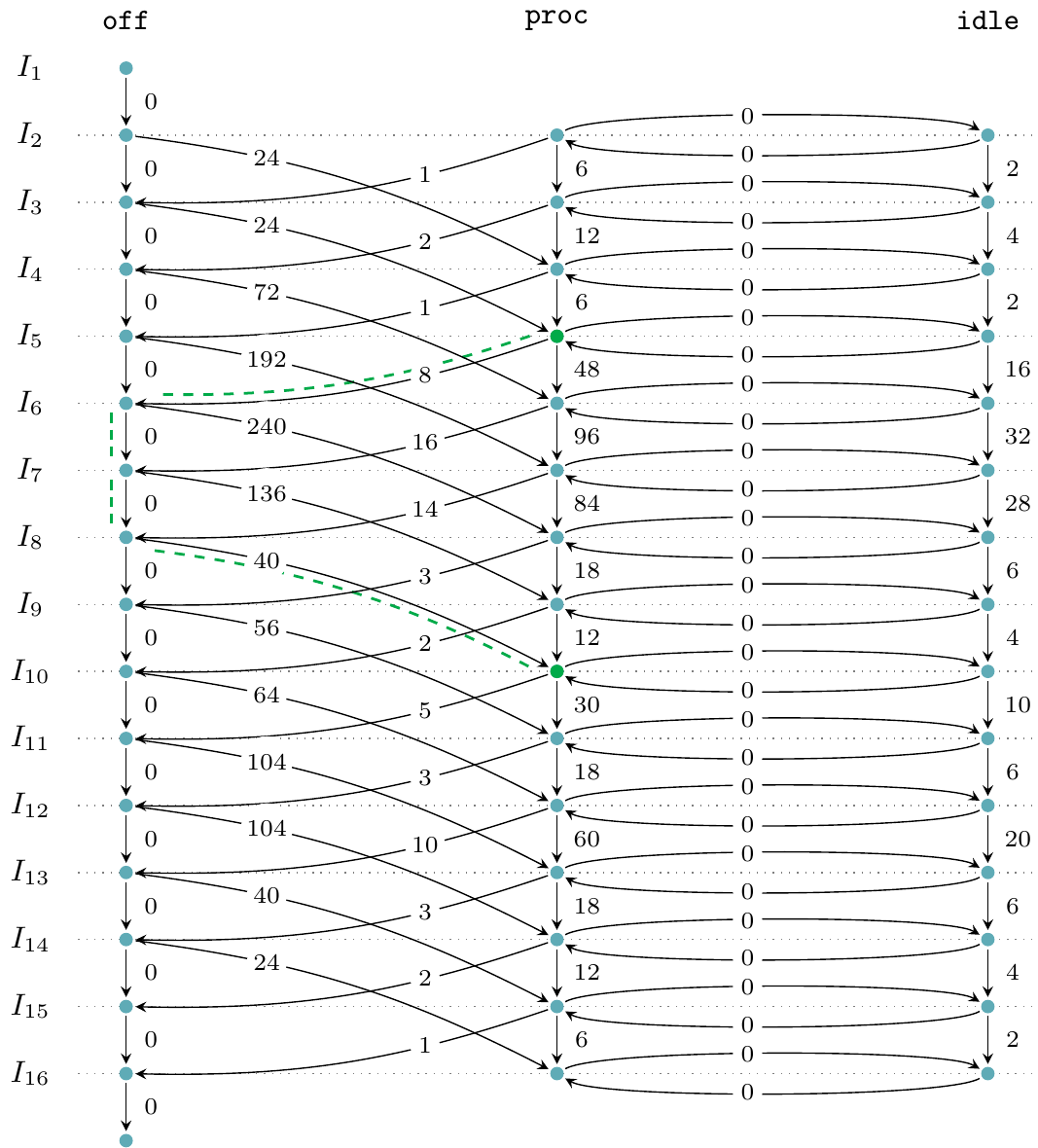}
    \caption{Interval-state graph for the Example instance from \cref{sec:problem-statement} with highlighted optimal switching behavior from $ (\StateProc, \Interval{4}) $ to $ (\StateProc, \Interval{10}) $.}
    \label{fig:states-graph}
\end{figure}

\paragraph{Example (continued):} Continuing the Example, \cref{fig:states-graph} shows the whole interval-state graph for the given instance.  
The green dashed path shows the optimal switching behavior of the machine assuming that the machine is in $\StateProc$ state during intervals $\Interval{4}$ and $\Interval{10}$; at first the machine is turned off (during $\Interval{5}$), then it remains off (during intervals $\Interval{6}$ and $\Interval{7}$), and is turned on afterward (intervals $\Interval{8}$, $\Interval{9}$). The optimal switching cost $\OptCostTrans[4,10]$ is, in this case, 48. The optimal switching behavior is 
\begin{equation}
    \OptPathTrans[4,10] = ((\StateProc, \StateOff), (\StateOff, \StateOff), (\StateOff, \StateOff), (\StateOff,\StateProc), (\StateOff,\StateProc))\,.
\end{equation}

\subsection{Integer Linear Programming Model \AbbrIlpOur{}}
In the \AbbrILP{} model proposed in \cite{2017:Aghelinejad}, the state transition functions are explicitly encoded.
In contrast, our \AbbrIlpOur{} model works only with the optimal switching costs pre-computed by the \AbbrIwa{} algorithm, thus encoding the transitions implicitly without sacrificing the optimality.
The only task of the \AbbrILP{} solver is then to schedule the jobs, and select appropriate spaces in between, such that the TEC is minimized.
Thus, the structure of our model is greatly simplified, with positive impact on its performance.

Formally, the variables used in the \AbbrIlpOur{} model are
\begin{itemize}
    \item job start time $\VarJobStart{\IdxJob}{\IdxInterval} \in \{0, 1\}$: equals 1 if job $\Job{\IdxJob}$ starts at the beginning of interval $\Interval{\IdxInterval}$, otherwise 0; \\
    \item space activation $\VarGap{\IdxInterval}{\IdxAnother{\IdxInterval}} \in \{0 ,1\}$: equals 1 if the machine undergoes the optimal switching defined by \hbox{$\OptPathTrans[\IdxInterval,\IdxAnother{\IdxInterval}]$}, otherwise 0.
\end{itemize}

The complete model follows.
\begin{align}
    &   \min \sum\limits_{\substack{\Interval{\IdxInterval}, \Interval{\IdxAnother{\IdxInterval}} \in \SetIntervals \\ \IdxInterval < \IdxAnother{\IdxInterval}}} \VarGap{\IdxInterval}{\IdxAnother{\IdxInterval}} \cdot \OptCostTrans[\IdxInterval,\IdxAnother{\IdxInterval}] 
    + \sum\limits_{\substack{\Job{\IdxJob} \in \SetJobs \\ \Interval{\IdxInterval} \in \SetIntervals }} \VarJobStart{\IdxJob}{\IdxInterval} \cdot \JobCost{\IdxJob}{\IdxInterval},  \label{eq:ilpmodel-objective}\\
     & \sum\limits_{\Interval{\IdxInterval} \in \SetIntervals} \VarJobStart{\IdxJob}{\IdxInterval} = 1, \  \forall \Job{\IdxJob} \in \SetJobs, \label{eq:model-schedule-every-job} \\ 
     & \VarJobStart{\IdxJob}{\IdxInterval} = 0, \ \forall \Job{\IdxJob} \in \SetJobs, \forall \IdxInterval \in \IntInterval{1}{\FirstOn-1} \cup \IntInterval{\LastOn - \ProcTime{\IdxJob} + 2}{\NumIntervals}, \label{eq:model-start-out-of-bound} \\
    & \sum\limits_{\Job{\IdxJob} \in \SetJobs}
    \sum_{\IdxAnother{\IdxInterval} = \max\{2,\IdxInterval - \ProcTime{\IdxJob} + 1\}}^{\IdxInterval} \VarJobStart{\IdxJob}{\IdxAnother{\IdxInterval}}
    + \sum_{\IdxAnother{\IdxInterval} = 1}^{\IdxInterval-1} \sum\limits_{\IdxAnotherTwo{\IdxInterval} = \IdxInterval+1}^{\NumIntervals} \VarGap{\IdxAnother{\IdxInterval}}{\IdxAnotherTwo{\IdxInterval}}
    = 1, \, \forall \Interval{\IdxInterval} \in \{\Interval{2}, \Interval{3}, \dots, \Interval{\NumIntervals-1}\}. \label{eq:model-no-overlap}
\end{align}
The objective~\eqref{eq:ilpmodel-objective} minimizes the total energy cost, consisting of the optimal switching cost of active spaces, and the cost of jobs processing, where
\begin{equation}
    \JobCost{\IdxJob}{\IdxInterval} = \sum_{\IdxAnother{\IdxInterval} = \IdxInterval}^{\IdxInterval + \ProcTime{\IdxJob} - 1} \EnergyCost{\IdxAnother{\IdxInterval}} \cdot \TransPower[\StateProc, \StateProc]
\end{equation}
for job $\Job{\IdxJob} \in \SetJobs$ and $\IdxInterval \in \IntInterval{\FirstOn}{\LastOn - \ProcTime{\IdxJob} + 1}$.

Constraint~\eqref{eq:model-schedule-every-job} forces every job to be scheduled exactly once, and constraint~\eqref{eq:model-start-out-of-bound} forbids the job to be scheduled before $\Interval{\FirstOn}$ and after $\Interval{\LastOn}$.
Finally, the last constraints~\eqref{eq:model-no-overlap} force the machine to be processing a job or to be undergoing some transition during every interval and forbid overlaps between them.

\subsubsection{Search Space Reduction}
Various methods can be employed to reduce the search space without sacrificing the optimality.
One of such methods is pruning of the spaces variables that lead to infeasible solutions if activated.

The pruning works as follows.
For each \( \Interval{\IdxInterval}, \Interval{\IdxAnother{\IdxInterval}} \) such that \( \IdxInterval <  \IdxAnother{\IdxInterval} \), the available time for processing the jobs is computed for both left (before \( \Interval{\IdxInterval} \)) and right (after \( \Interval{\IdxAnother{\IdxInterval}} \)) part of the scheduling horizon, i.e., \( \IdxInterval - \FirstOn + 1\) and \( \LastOn - \IdxAnother{\IdxInterval} + 1\), respectively.
Then, activating the switching behavior \( \OptPathTrans[\IdxInterval,\IdxAnother{\IdxInterval}] \) leads to an infeasible solution if one of the following \DefTerm{pruning conditions} holds
\begin{enumerate}[label={PC.\arabic*:},leftmargin=\widthof{[PC.2]}+\labelsep]
    \item The largest job can be fitted in neither part, i.e.,
    \begin{equation}
        \max_{\Job{\IdxJob} \in \SetJobs} \ProcTime{\IdxJob} > \IdxInterval - \FirstOn + 1 \quad \wedge \quad \max_{\Job{\IdxJob} \in \SetJobs} \ProcTime{\IdxJob} > \LastOn - \IdxAnother{\IdxInterval} + 1\,.
    \end{equation}
    \item The total available time for processing is less than the sum of all the processing times, i.e.,
    \begin{equation}
        (\IdxInterval - \FirstOn + 1) + (\LastOn - \IdxAnother{\IdxInterval} + 1) < \sum_{\Job{\IdxJob} \in \SetJobs} \ProcTime{\IdxJob}\,.
    \end{equation}
\end{enumerate}
If any of these conditions holds, the corresponding space variable \( \VarGap{\IdxInterval}{\IdxAnother{\IdxInterval}} \) is not created in \AbbrIlpOur{}.

\subsection{Constraint Programming Model \AbbrCpOur{}}

The idea of the \AbbrCpOur{} model is similar to the \AbbrIlpOur{}, with the exception that the spaces are not fixed -- they are allowed to `float' within the scheduling horizon. In consequence, the spaces do not have fixed costs because the cost depends on the position of the space in the horizon and its length.
In our \AbbrCpOur{} model, costs are formulated with an \textit{Element} expression, which is integrated in the objective.
To describe the \AbbrCpOur{} model, we use the IBM CP formalism \cite{2018:Laborie}.

\paragraph{Variables:} Three types of interval variables are used in the \AbbrCP{} model. 

To represent the jobs, we use optional interval variables  $\CpVarJobAlt{\IdxJob}{\IdxInterval}$, $\forall \Job{\IdxJob} \in \SetJobs, \IdxInterval \in \IntInterval{ \FirstOn}{\LastOn - \ProcTime{\IdxJob} + 1}$, which model whether job $\Job{\IdxJob}$ starts at the beginning of the interval $\Interval{\IdxInterval}$.
Only one such variable, represented by interval variable $\CpVarJob{\IdxJob}$, will be present in the schedule for each job. Length of $\CpVarJobAlt{\IdxJob}{\IdxInterval}$ is fixed to $\ProcTime{\IdxJob}$, and its start is fixed to $\IdxInterval - 1$.

Finally, the optional interval variables $\CpVarGap{\IdxGapLen}{\IdxGapInc}$ represent the `floating' spaces of fixed length.
For each possible \emph{length} $\IdxGapLen \in \{ 1, 2, \dots, \NumIntervals - 2 - \sum_{\Job{\IdxJob} \in \SetJobs} \ProcTime{\IdxJob}\}$, we create $\GapMaxInc{\IdxGapLen} = \left\lfloor \frac{ \NumIntervals - 2 - \sum_{\Job{\IdxJob} \in \SetJobs} \ProcTime{\IdxJob}}{\IdxGapLen} \right\rfloor$ variables that are indexed by $\IdxGapInc \in \{ 1, 2 \dots, \GapMaxInc{\IdxGapLen} \}$. Note that the number $\GapMaxInc{\IdxGapLen}$ gives the upper bound on the number of the spaces of length $\IdxGapLen$ that may appear in a feasible schedule, while $\GapMaxLen = \NumIntervals - 2 - \sum_{\Job{\IdxJob} \in \SetJobs} \ProcTime{\IdxJob}$ gives an upper bound on the space length.

\paragraph{Constraints:} Since the machine is assumed to be in $\StateOff$ state during $\Interval{1}$ and $\Interval{\NumIntervals}$, the earliest and the latest interval during which a switching might occur is $\Interval{2}$ and $\Interval{\NumIntervals - 1}$, respectively. Hence, starts (ends) of the spaces are restricted by

\begin{equation}
\left.
\begin{aligned}
    \CpStartOf{\CpVarGap{\IdxGapLen}{\IdxGapInc}} & \geq 1 \\
    \CpEndOf{\CpVarGap{\IdxGapLen}{\IdxGapInc}} & \leq \NumIntervals-1
\end{aligned} \right\rbrace  \ \forall \IdxGapLen \in \IntInterval{1}{\GapMaxLen}, \IdxGapInc \in \IntInterval{1}{\GapMaxInc{\IdxGapLen}}.
\end{equation}
As mentioned previously, the spaces have fixed lengths, i.e.,
\begin{equation}
    \CpLengthOf{\VarGap{\IdxGapLen}{\IdxGapInc}} = \IdxGapLen, \  \forall \IdxGapLen \in \IntInterval{1}{\GapMaxLen}, \IdxGapInc \in \IntInterval{1}{\GapMaxInc{\IdxGapLen}}.
\end{equation}
Relationship between variables $\CpVarJob{\IdxJob}$ and $\CpVarJobAlt{\IdxJob}{\IdxInterval}$ is given by
\begin{equation}
    \CpAlternative{\CpVarJob{\IdxJob}, \{\CpVarJobAlt{\IdxJob}{\IdxInterval} : \IdxInterval \in \IntInterval{\FirstOn}{\LastOn - \ProcTime{\IdxJob} + 1}\}}, \ \forall \Job{\IdxJob} \in \SetJobs,
\end{equation}
To ensure that jobs and spaces are not overlapping, we use the \textit{NoOverlap} constraint,
\begin{equation}
     \CpNoOverlap{\{ \CpVarGap{\IdxGapLen}{\IdxGapInc} : \IdxGapLen \in \IntInterval{1}{\GapMaxLen}, \IdxGapInc \in \IntInterval{1}{\GapMaxInc{\IdxGapLen}} \} \cup \{\CpVarJob{\IdxJob} : \Job{\IdxJob} \in \SetJobs\}}.
\end{equation}
The lengths of the spaces are constrained by
\begin{equation}
    \sum_{\IdxGapLen=1}^{\GapMaxLen} \sum_{\IdxGapInc=1}^{\GapMaxInc{\IdxGapLen}} \CpLengthOf{\CpVarGap{\IdxGapLen}{\IdxGapInc}} = \GapMaxLen,
\end{equation}
to ensure that the whole scheduling horizon is filled.

Finally, to eliminate some symmetries in the model, space variables $\CpVarGap{\IdxGapLen}{\IdxGapInc}$ are constrained such that space with index $\IdxGapInc$ can be present in the solution only if all the spaces $\CpVarGap{\IdxGapLen}{\IdxAnother{\IdxGapInc}}$ of the same length with $\IdxAnother{\IdxGapInc} < \IdxGapInc$ are present, i.e.,

\begin{equation}
    \CpPresenceOf{\CpVarGap{\IdxGapLen}{\IdxGapInc}} \geq \CpPresenceOf{\CpVarGap{\IdxGapLen}{\IdxGapInc+1}}, \ \forall \IdxGapLen \in \IntInterval{1}{\GapMaxLen}, \IdxGapInc \in \IntInterval{1}{\GapMaxInc{\IdxGapLen} - 1}.
\end{equation}

\paragraph{Objective:} The objective is to minimize the TEC, here expressed as 
\begin{equation}
    \sum_{\IdxGapLen=1}^{\GapMaxLen} \sum_{\IdxGapInc=1}^{\GapMaxInc{\IdxGapLen}} \CpElement{\CpElementArray{\IdxGapLen}, \CpStartOf{\CpVarGap{\IdxGapLen}{\IdxGapInc}}}
    + \sum_{\Job{\IdxJob} \in \SetJobs} \sum_{\IdxInterval=\FirstOn}^{\LastOn - \ProcTime{\IdxJob} + 1} \CpPresenceOf{\CpVarJobAlt{\IdxJob}{\IdxInterval}} \cdot \JobCost{\IdxJob}{\IdxInterval},
\end{equation}

where the first part corresponds to the cost for optimal switchings between the job processings, and the second part corresponds to the cost for job processing.
To compute the cost of the present spaces, vector
\begin{equation}
    \CpElementArray{\IdxGapLen} = (\OptCostTrans[1, 1 + \IdxGapLen + 1], \OptCostTrans[2, 2 + \IdxGapLen + 1], \dots, \OptCostTrans[\NumIntervals - \IdxGapLen - 1, \NumIntervals])
\end{equation}
is used to represent the optimal switching costs for the given $\IdxGapLen$ addressed by the start of space $\CpVarGap{\IdxGapLen}{\IdxGapInc}$ (indexed from 1).

\begin{remark}
Note that there are many different ways of implementing the \AbbrCP{} model.
For example, each possible space might be fixed in time, similarly as in \AbbrIlpOur{}. Then, the objective would simplify to a sum of spaces presences multiplied by their costs.
Another alternative would be to have `floating' spaces with variable lengths. In that case, the number of the interval variables needed to represent the spaces would decrease to $\NumJobs - 1$, but the element expression in the objective would need to be indexed by both the start time and the length of each space.
Also, the element expressions could be replaced by overlap expressions, etc.
We have tried multiple different alternatives; however, the performances of the models on preliminary benchmark instances were more or less similar. The described \AbbrCP{} model was slightly better than the others; therefore, we use it for the experiments in \cref{sec:experiments}.
\end{remark}

\begin{remark}
We can use a similar idea to reduce the search space as applied to \AbbrIlpOur{}.
However, in the case of \AbbrCpOur{}, the spaces variables cannot be pruned since they are not fixed in time.
Instead, we can enforce the corresponding cost of this space to a large number, which effectively deactivates it.
\end{remark}

\section{Experiments} \label{sec:experiments}

\setlength{\tabcolsep}{5pt}

This section evaluates how \AbbrIlpOur{} and \AbbrCpOur{} models perform in comparison to the \AbbrIlpRef{} model proposed in \cite{2017:Aghelinejad}.
The comparison is made on a set of randomly generated instances with varying sizes and different machine transition graphs; see \cref{sec:instances} for the description of the generated dataset.
The results are presented in \cref{sec:experiment-results}.

All experiments were executed on 2x Intel(R) Xeon(R) Silver 4110 CPU \@ \SI{2.10}{\giga\hertz} with \SI{188}{\giga\byte} of RAM (16 cores in total).
For solving the \AbbrILP{} and \AbbrCP{} models, we used Gurobi 8 and IBM CP Optimizer 12.9, respectively.
Except for the time-limit and search phases in \AbbrCpOur{}, which branched on jobs first, all the solver parameters were set to default values.

The generated instances and their solutions are publicly available at \\ \url{https://github.com/CTU-IIG/EnergyStatesAndCostsSchedulingDatasets}.

\subsection{Instances}\label{sec:instances}
\begin{figure}[t]
    \centering
	\includegraphics{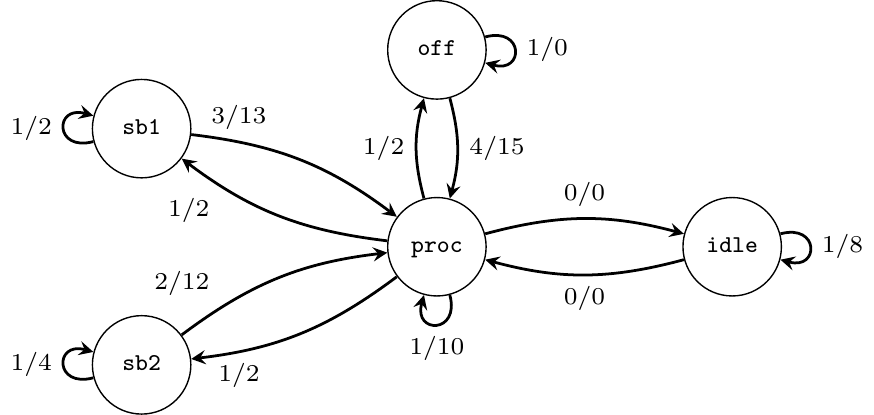}
    \caption{Example of a transition graph with multiple standby states; every edge from $\IdxState$ to $\IdxAnother{\IdxState}$ is labeled by $\TransTime[\IdxState, \IdxAnother{\IdxState}]$/$\TransPower[\IdxState, \IdxAnother{\IdxState}]$.}
    \label{fig:machine-multioff}
\end{figure}

The instances in the dataset can be divided according to
\begin{enumerate}
    \item a number of jobs:
        \begin{enumerate}
            \item \DataMedium{}: medium instances with \( \NumJobs \in \{ 30, 60, 90 \} \);
            \item \DataLarge{}: large instances with \( \NumJobs \in \{ 150, 170, 190 \} \);
        \end{enumerate}

    \item a machine transition graph:
        \begin{enumerate}
            \item \DataSingleOff{}: a simple model with no standby state used by~\cite{2014:Shrouf,2017:Aghelinejad}, see \cref{fig:example-func-power-time} for its description;
            \item \DataMultiSby{}: a model with two standby states shown in \cref{fig:machine-multioff}.
        \end{enumerate}
\end{enumerate}
For fixed \( \NumJobs \) and a machine transition graph, 12 random instances are generated in the following way (48 instances in the whole dataset).
The processing times of the jobs are randomly sampled from discrete uniform distribution \(\UnifDistDiscrete{1}{5} \).
The number of intervals in each instance is obtained as a multiple of the total processing time plus the required number of intervals to turn the machine on and off, where the multiple is taken from set the $ \{1.3, 1.6, 1.9, 2.2\}$.
The energy cost in each interval is randomly sampled from \(\UnifDistDiscrete{1}{10} \).
For instances differing only in the number of intervals, the energy costs are sampled gradually, i.e., the energy costs of all the intervals in an instance with a shorter horizon are the same as for the corresponding intervals in an instance with a longer horizon.

Note that the distributions for sampling the processing times and the energy costs of the intervals are the same as proposed in~\cite{2014:Shrouf,2017:Aghelinejad}.

\subsection{Results}\label{sec:experiment-results}

All the presented Tables \ref{tab:experiment-medium-singleoff}, \ref{tab:experiment-medium-multioff},  \ref{tab:experiment-large-singleoff}, and \ref{tab:experiment-large-multioff} have the same structure: each row represents one instance characterized by the number of the jobs \( \NumJobs \) and the number of intervals \( \NumIntervals \).
The objective value \( \AbbrObjective{} \) of the found feasible solution, lower bound \( \AbbrLb{} \) and the running time \( \AbbrTime \) are given for each tested model.
If the objective value or the lower bound is in bold font, the corresponding value is known to be optimal.
Therefore, if both objective and the lower bound are in bold, the solver was able to prove the solution optimality within the time-limit.
If the solver reached its given time-limit on an instance without proving the optimality of a solution, the value in the corresponding cell in \( \AbbrTime \) column is \AbbrTimeLimit{}.

Additionally, we report the pre-processing time \AbbrPreProc{} for the large instances. For medium-size instances, the pre-processing time is negligible with the average time \Second{0.69} and maximum time \Second{2.93}.

The last rows in each table shows the average running time on each model and the average optimality gap.
The average time is computed over all instances; if the solver time-outed on some instance, the specified time-limit is taken as the running time on that instance.
The optimality gap on each instance is defined as
\begin{equation}
    \frac{\AbbrObjective{} - \AbbrLbBest}{\AbbrObjective{}} \cdot 100 \  [\si{\percent}]\,,
\end{equation}
where \( \AbbrLbBest{} \) is the best lower bound obtained over all models on that instance.

\subsubsection{Results for Medium Instances}\label{sec:experiment-medium}
The results of the experiment with medium-size instances for \DataSingleOff{} and \DataMultiSby{} transition graphs are shown in \cref{tab:experiment-medium-singleoff} and \cref{tab:experiment-medium-multioff}, respectively.
In these tables we can see that \AbbrIlpOur{} finds the optimal solutions and proves their optimality for all instances.
On the other hand, the model \AbbrIlpRef{} proposed in~\cite{2017:Aghelinejad} finds the optimal solution and proves the optimality only for 11 instances out of 24 within the time-limit (\Second{600}).
Moreover, some of the non-optimal solutions found by \AbbrIlpRef{} are far from the optimum, for example, the objective of the solution found for \( \NumJobs = 90, \NumIntervals = 621 \) on \DataMultiSby{} is more than twice the objective of the optimal one found by \AbbrIlpOur{}.

Unfortunately, \AbbrCpOur{} is not able to prove the optimality of any instance within the time-limit.
However, the average optimality gaps (\Percent{3.40} for \DataSingleOff{} and \Percent{1.79} for \DataMultiSby{}) reveals that it can find near-optimal solutions.
The performance of both \AbbrCpOur{} and \AbbrIlpOur{} is slightly influenced by a more complex transition graph, whereas the performance of \AbbrIlpRef{} deteriorates significantly (average optimality gap \Percent{2.14} for \DataSingleOff{} increased to \Percent{29.59} for \DataMultiSby{}).

\begin{table}[H]
    \centering
    \caption{\DataMedium{}+\DataSingleOff{}: Comparison of found upper bound \AbbrObjective{}, lower bound \AbbrLb{} and runtime \AbbrTime{} between the models. Time-limit is \Second{600} and \AbbrTimeLimit{} stands for time-limit reached.} \label{tab:experiment-medium-singleoff}
    \input{tables/experiment-medium-singleoff.tex}
\end{table}
    
\begin{table}[H]
    \centering
    \caption{\DataMedium{}+\DataMultiSby{}: Comparison of found upper bound \AbbrObjective{}, lower bound \AbbrLb{} and runtime \AbbrTime{} between the models. Time-limit is \Second{600} and \AbbrTimeLimit{} stands for time-limit reached.} \label{tab:experiment-medium-multioff}
    \input{tables/experiment-medium-multioff.tex}
\end{table}

\clearpage
\subsubsection{Results for Large Instances}\label{sec:experiment-large}
The results of the experiment with large instances for \DataSingleOff{} and \DataMultiSby{} transition graphs are shown in \cref{tab:experiment-large-singleoff} and \cref{tab:experiment-large-multioff}, respectively. The results for \AbbrCpOur{} are not included, since we were unable to obtain solutions to all the instances from the IBM CP Optimizer. We observed that the solver used all the available RAM and swap memory ($\SI{188}{\giga\byte}+\SI{191}{\giga\byte}$), which indicates internal issues of the solver. However, for three instances where the \AbbrCP{} solver was able to find the solutions, the objective was better than for \AbbrIlpRef{}.

Looking at the results of \AbbrIlpOur{}, we can see that it found the optimal solutions and proved their optimality for all 24 instances. On the other hand, \AbbrIlpRef{} was able to find the optimal solutions for only two smallest instances, but was not able to prove their optimality within the specified time-limit (\Second{3600}).
Comparing the average optimality gaps, \AbbrIlpRef{} achieved \Percent{8.39} on \DataSingleOff{} transition graph and \Percent{61.33} on \DataMultiSby{}, whereas \AbbrIlpOur{} achieved \Percent{0} optimality gap on both transition graphs.
This shows that \AbbrIlpOur{} scales even to large instances.

\begin{table}[H]
    \centering
    \caption{\DataLarge{}+\DataSingleOff{}: Comparison of found upper bound \AbbrObjective{}, lower bound \AbbrLb{} and runtime \AbbrTime{} between the models. Time-limit is \Second{3600} and \AbbrTimeLimit{} stands for time-limit reached.} \label{tab:experiment-large-singleoff}
    \input{tables/experiment-large-singleoff.tex}
\end{table}

\clearpage
\section{Conclusions} \label{sec:conclusions}

Continuing on the recent research of the single-machine scheduling problem with the variable energy costs and power-saving machine states, we propose a pre-processing algorithm \AbbrIwa{}, which pre-computes the optimal switching behavior of the machine for all possible spaces in the schedule.
The pre-processing can be done in polynomial time and works well even for large instances of the problem, e.g., it takes \Second{23} to pre-process our largest benchmark instance with 190 jobs and 1277 intervals.
The pre-computed switching costs are successfully integrated into novel \AbbrCP{} and \AbbrILP{} models, which are compared to the state-of-the-art exact \AbbrILP{} model on a set of benchmark instances.
Results show that our approach outperforms the existing methods considering all aspects -- the runtime, the provided lower bounds and the upper bounds.
Using our models, we obtain the optimal solutions even for the large instances with up to 190 jobs and 1277 intervals, which have been previously tackled only heuristically~\cite{2017:Aghelinejad}.

\begin{table}[t]
    \centering
    \caption{\DataLarge{}+\DataMultiSby{}: Comparison of found upper bound \AbbrObjective{}, lower bound \AbbrLb{} and runtime \AbbrTime{} between the models. Time-limit is \Second{3600} and \AbbrTimeLimit{} stands for time-limit reached.} \label{tab:experiment-large-multioff}
    \input{tables/experiment-large-multioff.tex}
\end{table}

\clearpage
%
%
\bibliographystyle{splncs04}
\bibliography{references}

\end{document}

%% file: macros.tex
\newcommand{\AbbrIwa}{SPACES}
\newcommand{\AbbrIlpRef}{ILP-REF}
\newcommand{\AbbrIlpOur}{ILP-SPACES}
\newcommand{\AbbrCpOur}{CP-SPACES}
\newcommand{\AbbrPreProc}{P-P}
\newcommand{\AbbrObjective}{\textit{ub}}
\newcommand{\AbbrLb}{\textit{lb}}
\newcommand{\AbbrLbBest}{{\textit{lb}}^{\text{best}}}
\newcommand{\AbbrTime}{\textit{t}}
\newcommand{\AbbrTimeLimit}{TLR}
\newcommand{\AbbrILP}{ILP}
\newcommand{\AbbrCP}{CP}

\newcommand{\DataMedium}{MEDIUM}
\newcommand{\DataLarge}{LARGE}
\newcommand{\DataSingleOff}{NOSBY}
\newcommand{\DataMultiSby}{TWOSBY}

\newcommand{\Percent}[1]{\SI{#1}{\percent}}
\newcommand{\Second}[1]{\SI{#1}{\second}}

\newcommand{\Optimum}[1]{\bfseries #1}
\newcommand{\TabColSep}{0.7cm}

\newcommand{\DefTerm}[1]{\emph{#1}} 
\newcommand{\AsymCompUpper}[1]{\mathcal{O}(#1)}
\newcommand{\UnifDistDiscrete}[2]{\mathcal{U}\{#1,#2\}}  

\newcommand{\SetJobs}{\mathcal{J}} 
\newcommand{\SetIntervals}{\mathcal{I}} 
\newcommand{\SetStates}{\mathcal{S}} 
\newcommand{\SetIntNonNeg}{\mathbb{Z}_{\geq 0}} 
\newcommand{\SetIntPos}{\mathbb{Z}_{> 0}} 
\newcommand{\SetVertTrans}{V} 
\newcommand{\SetEdgesTrans}{E} 

\newcommand{\Job}[1]{J_{#1}} 
\newcommand{\Interval}[1]{I_{#1}} 

\newcommand{\IdxJob}{j}
\newcommand{\IdxInterval}{i}
\newcommand{\IdxState}{s}
\newcommand{\IdxAnother}[1]{#1^{\prime}} %
\newcommand{\IdxAnotherTwo}[1]{#1^{\prime\prime}} %

\newcommand{\NumJobs}{n}
\newcommand{\NumIntervals}{h}
\newcommand{\NumStates}{\vert \SetStates \vert}

\newcommand{\ProcTime}[1]{p_{#1}}
\newcommand{\EnergyCost}[1]{c_{#1}}
\newcommand{\EnergyCostVector}{\bm{c}}

\newcommand{\TransTimeSymbol}{T}
\DeclareDocumentCommand \TransTime {o} {
  \IfNoValueTF{#1} {
        \TransTimeSymbol
  }{
        \TransTimeSymbol(#1)
  }
}

\newcommand{\TransPowerSymbol}{P}
\DeclareDocumentCommand \TransPower {o} {
  \IfNoValueTF{#1} {
        \TransPowerSymbol
  }{
        \TransPowerSymbol(#1)
  }
}

\newcommand{\ProblemNotation}{1,\, \text{TOU} \vert \, \text{states} \, \vert \, \text{TEC}}

\newcommand{\StateOff}{\texttt{off}}
\newcommand{\StateIdle}{\texttt{idle}}
\newcommand{\StateProc}{\texttt{proc}}

\newcommand{\SolVecStarts}{\bm{\sigma}}
\newcommand{\SolStarts}[1]{\sigma_{#1}}
\newcommand{\SolVecTrans}{\boldsymbol{\Omega}}
\newcommand{\SolTrans}[1]{\Omega_{#1}}

\newcommand{\IntInterval}[2]{\{#1, \dots, #2\}}

\newcommand{\VertTrans}[2]{v_{#1,#2}}   

\newcommand{\WeightTransSymbol}{w}
\DeclareDocumentCommand \WeightTrans {o} {
  \IfNoValueTF{#1} {
        \WeightTransSymbol
  }{
        \WeightTransSymbol(#1)
  }
}

\newcommand{\PathCostTransSymbol}{l}
\DeclareDocumentCommand \PathCostTrans {o} {
  \IfNoValueTF{#1} {
        \PathCostTransSymbol
  }{
        \PathCostTransSymbol(#1)
  }
}

\newcommand{\OptCostTransSymbol}{c^{\star}}
\DeclareDocumentCommand \OptCostTrans {o} {
  \IfNoValueTF{#1} {
        \OptCostTransSymbol
  }{
        \OptCostTransSymbol(#1)
  }
}

\newcommand{\OptPathTransSymbol}{\SolVecTrans^{\star}}
\DeclareDocumentCommand \OptPathTrans {o} {
  \IfNoValueTF{#1} {
        \OptPathTransSymbol
  }{
        \OptPathTransSymbol(#1)
  }
}

\newcommand{\FirstOn}{{\text{earl}}} 
\newcommand{\LastOn}{{\text{late}}} 

\newcommand{\JobCost}[2]{c^{\text{(job)}}_{#1, #2}} 

\newcommand{\VarJobStart}[2]{s_{#1,#2}} 
\newcommand{\VarGap}[2]{x_{#1,#2}} 

\SetKwProg{Fn}{Function}{:}{}
\SetKwFunction{KwIwa}{\AbbrIwa{}}
\SetKw{Continue}{continue}
\SetKw{Break}{break}
\SetKw{True}{TRUE}
\SetKw{False}{FALSE}

\DeclareMathOperator{\CmdCpStartOf}{StartOf}
\DeclareMathOperator{\CmdCpEndOf}{EndOf}
\DeclareMathOperator{\CmdCpLengthOf}{LengthOf}
\DeclareMathOperator{\CmdCpPresenceOf}{PresenceOf}
\DeclareMathOperator{\CmdCpAlternative}{Alternative}
\DeclareMathOperator{\CmdCpElement}{Element}
\DeclareMathOperator{\CmdCpNoOverlap}{NoOverlap}

\newcommand{\CpLengthOf}[1]{\CmdCpLengthOf(#1)}
\newcommand{\CpStartOf}[1]{\CmdCpStartOf(#1)}
\newcommand{\CpEndOf}[1]{\CmdCpEndOf(#1)}
\newcommand{\CpAlternative}[1]{\CmdCpAlternative(#1)}
\newcommand{\CpElement}[1]{\CmdCpElement(#1)}
\newcommand{\CpPresenceOf}[1]{\CmdCpPresenceOf(#1)}
\newcommand{\CpNoOverlap}[1]{\CmdCpNoOverlap(#1)}

\newcommand{\CpVarJobAlt}[2]{s_{#1, #2}}
\newcommand{\CpVarJob}[1]{s_{#1}}
\newcommand{\CpVarGap}[2]{\VarGap{#1}{#2}}
\newcommand{\IdxGapLen}{\ell}
\newcommand{\IdxGapInc}{k}
\newcommand{\GapMaxInc}[1]{K(#1)}
\newcommand{\GapMaxLen}{{\ell}_{\text{max}}}
\newcommand{\CpElementArray}[1]{\bm{c}^{\star}_{#1}}

%% file: tables/experiment-medium-singleoff.tex
\begin{adjustbox}{max width=\textwidth}

{
\renewrobustcmd{\bfseries}{\fontseries{b}\selectfont}
\begin{tabular}{
    S[table-format=2.0]
    S[table-format=3.0] @{\hskip \TabColSep}
    S[table-format=4.0]
    S[table-format=4.0]
    S[table-format=2.1] @{\hskip \TabColSep}
    S[table-format=4.0]
    S[table-format=4.0]
    S[table-format=2.1] @{\hskip \TabColSep}
    S[table-format=4.0]
    S[table-format=4.0]
    S[table-format=3.1]}
    \toprule
    \multicolumn{2}{c @{\hskip \TabColSep}}{Instance} & 
    \multicolumn{3}{c @{\hskip \TabColSep}}{\AbbrIlpRef{}~\cite{2017:Aghelinejad}} &
    \multicolumn{3}{c @{\hskip \TabColSep}}{\AbbrCpOur{}} &
    \multicolumn{3}{c}{\AbbrIlpOur{}} \\
    \cmidrule(lr{\TabColSep}){1-2} \cmidrule(lr{\TabColSep}){3-5} \cmidrule(lr{\TabColSep}){6-8} \cmidrule{9-11}
    \multicolumn{1}{c}{\scriptsize $\NumJobs$} &
    \multicolumn{1}{c @{\hskip \TabColSep}}{\scriptsize $\NumIntervals$} &
    \multicolumn{1}{c}{\scriptsize \AbbrObjective{} [-]} & \multicolumn{1}{c}{\scriptsize \AbbrLb{} [-]} &
    \multicolumn{1}{c @{\hskip \TabColSep}}{\scriptsize \AbbrTime{} [\si{\second}]} &
    \multicolumn{1}{c}{\scriptsize \AbbrObjective{} [-]} & \multicolumn{1}{c}{\scriptsize \AbbrLb{} [-]} &
    \multicolumn{1}{c @{\hskip \TabColSep}}{\scriptsize \AbbrTime{} [\si{\second}]} &
    \multicolumn{1}{c}{\scriptsize \AbbrObjective{} [-]} & \multicolumn{1}{c}{\scriptsize \AbbrLb{} [-]} & 
    \multicolumn{1}{c}{\scriptsize \AbbrTime{} [\si{\second}]} \\
    \midrule
    30 & 104 & \Optimum{1426} & \Optimum{1426} & 3.7 & 1448 & 496 & \AbbrTimeLimit{} & \Optimum{1426} & \Optimum{1426} & 1.3 \\
    30 & 127 & \Optimum{1394} & \Optimum{1394} & 4.9 & 1412 & 488 & \AbbrTimeLimit{} & \Optimum{1394} & \Optimum{1394} & 1.7 \\
    30 & 150 & \Optimum{1394} & \Optimum{1394} & 5.7 & 1414 & 484 & \AbbrTimeLimit{} & \Optimum{1394} & \Optimum{1394} & 2.4 \\
    30 & 173 & \Optimum{1394} & \Optimum{1394} & 7.0 & 1458 & 484 & \AbbrTimeLimit{} & \Optimum{1394} & \Optimum{1394} & 3.1 \\
    60 & 258 & \Optimum{4290} & \Optimum{4290} & 88.5 & 4379 & 1724 & \AbbrTimeLimit{} & \Optimum{4290} & \Optimum{4290} & 7.7 \\
    60 & 316 & \Optimum{3994} & \Optimum{3994} & 344.7 & 4117 & 1584 & \AbbrTimeLimit{} & \Optimum{3994} & \Optimum{3994} & 29.0 \\
    60 & 374 & \Optimum{3836} & 3826 & \AbbrTimeLimit{} & 3952 & 1424 & \AbbrTimeLimit{} & \Optimum{3836} & \Optimum{3836} & 29.0 \\
    60 & 432 & 3956 & 3800 & \AbbrTimeLimit{} & 3972 & 1380 & \AbbrTimeLimit{} & \Optimum{3833} & \Optimum{3833} & 46.9 \\
    90 & 363 & 6044 & 5839 & \AbbrTimeLimit{} & 6004 & 2328 & \AbbrTimeLimit{} & \Optimum{5920} & \Optimum{5920} & 7.0 \\
    90 & 445 & 5778 & 5567 & \AbbrTimeLimit{} & 5760 & 2232 & \AbbrTimeLimit{} & \Optimum{5686} & \Optimum{5686} & 166.0 \\
    90 & 528 & 5916 & 4695 & \AbbrTimeLimit{} & 5916 & 2168 & \AbbrTimeLimit{} & \Optimum{5431} & \Optimum{5431} & 64.5 \\
    90 & 610 & 5901 & 4514 & \AbbrTimeLimit{} & 5829 & 1828 & \AbbrTimeLimit{} & \Optimum{5373} & \Optimum{5373} & 147.1 \\
    \midrule
    \multicolumn{4}{l}{Average time [\si{\second}]:} & {337.9} & & & {\textgreater 600} & & & {42.1} \\
    \multicolumn{4}{l}{Average optimality gap [\si{\percent}]:} & {2.14} & & & {3.40} & & & {0.00} \\
    \bottomrule
    \end{tabular}
}

\end{adjustbox}

%% file: tables/experiment-medium-multioff.tex
\begin{adjustbox}{max width=\textwidth}

{
\renewrobustcmd{\bfseries}{\fontseries{b}\selectfont}
\begin{tabular}{
    S[table-format=2.0]
    S[table-format=3.0] @{\hskip \TabColSep}
    S[table-format=5.0]
    S[table-format=5.0]
    S[table-format=2.1] @{\hskip \TabColSep}
    S[table-format=5.0]
    S[table-format=4.0]
    S[table-format=2.1] @{\hskip \TabColSep}
    S[table-format=5.0]
    S[table-format=5.0]
    S[table-format=3.1]}
    \toprule
    \multicolumn{2}{c @{\hskip \TabColSep}}{Instance} & 
    \multicolumn{3}{c @{\hskip \TabColSep}}{\AbbrIlpRef{}~\cite{2017:Aghelinejad}} &
    \multicolumn{3}{c @{\hskip \TabColSep}}{\AbbrCpOur{}} &
    \multicolumn{3}{c}{\AbbrIlpOur{}} \\
    \cmidrule(lr{\TabColSep}){1-2} \cmidrule(lr{\TabColSep}){3-5} \cmidrule(lr{\TabColSep}){6-8} \cmidrule{9-11}
    \multicolumn{1}{c}{\scriptsize $\NumJobs$} &
    \multicolumn{1}{c @{\hskip \TabColSep}}{\scriptsize $\NumIntervals$} &
    \multicolumn{1}{c}{\scriptsize \AbbrObjective{} [-]} & \multicolumn{1}{c}{\scriptsize \AbbrLb{} [-]} &
    \multicolumn{1}{c @{\hskip \TabColSep}}{\scriptsize \AbbrTime{} [\si{\second}]} &
    \multicolumn{1}{c}{\scriptsize \AbbrObjective{} [-]} & \multicolumn{1}{c}{\scriptsize \AbbrLb{} [-]} &
    \multicolumn{1}{c @{\hskip \TabColSep}}{\scriptsize \AbbrTime{} [\si{\second}]} &
    \multicolumn{1}{c}{\scriptsize \AbbrObjective{} [-]} & \multicolumn{1}{c}{\scriptsize \AbbrLb{} [-]} & 
    \multicolumn{1}{c}{\scriptsize \AbbrTime{} [\si{\second}]} \\
    \midrule
    30 & 106 & \Optimum{3815} & \Optimum{3815} & 29.4 & \Optimum{3815} & 1240 & \AbbrTimeLimit{} & \Optimum{3815} & \Optimum{3815} & 1.4 \\
    30 & 129 & \Optimum{3804} & \Optimum{3804} & 30.7 & 3815 & 1220 & \AbbrTimeLimit{} & \Optimum{3804} & \Optimum{3804} & 2.3 \\
    30 & 152 & \Optimum{3804} & \Optimum{3804} & 42.0 & 3815 & 1210 & \AbbrTimeLimit{} & \Optimum{3804} & \Optimum{3804} & 7.0 \\
    30 & 175 & \Optimum{3804} & \Optimum{3804} & 61.4 & 3815 & 1210 & \AbbrTimeLimit{} & \Optimum{3804} & \Optimum{3804} & 9.5 \\
    60 & 254 & \Optimum{10863} & \Optimum{10863} & 588.1 & 11058 & 4190 & \AbbrTimeLimit{} & \Optimum{10863} & \Optimum{10863} & 2.0 \\
    60 & 311 & 10289 & 10087 & \AbbrTimeLimit{} & 10574 & 3860 & \AbbrTimeLimit{} & \Optimum{10248} & \Optimum{10248} & 43.3 \\
    60 & 368 & \Optimum{9917} & 9696 & \AbbrTimeLimit{} & 10163 & 3470 & \AbbrTimeLimit{} & \Optimum{9917} & \Optimum{9917} & 82.1 \\
    60 & 426 & 20346 & 9133 & \AbbrTimeLimit{} & 10055 & 3340 & \AbbrTimeLimit{} & \Optimum{9874} & \Optimum{9874} & 233.9 \\
    90 & 370 & 17179 & 14818 & \AbbrTimeLimit{} & 15470 & 5900 & \AbbrTimeLimit{} & \Optimum{15379} & \Optimum{15379} & 140.2 \\
    90 & 454 & 22808 & 12951 & \AbbrTimeLimit{} & 15156 & 5680 & \AbbrTimeLimit{} & \Optimum{14923} & \Optimum{14923} & 138.6 \\
    90 & 538 & 25992 & 11868 & \AbbrTimeLimit{} & 15107 & 5500 & \AbbrTimeLimit{} & \Optimum{14548} & \Optimum{14548} & 403.8 \\
    90 & 621 & 29558 & 11406 & \AbbrTimeLimit{} & 15152 & 4620 & \AbbrTimeLimit{} & \Optimum{14392} & \Optimum{14392} & 225.8 \\
    \midrule
    \multicolumn{4}{l}{Average time [\si{\second}]:} & {412.6} & & & {\textgreater 600} & & & {107.5} \\
    \multicolumn{4}{l}{Average optimality gap [\si{\percent}]:} & {29.59} & & & {1.79} & & & {0.00} \\
    \bottomrule
    \end{tabular}
}

\end{adjustbox}

%% file: tables/experiment-large-singleoff.tex
\begin{adjustbox}{max width=\textwidth}

\renewrobustcmd{\bfseries}{\fontseries{b}\selectfont}
    
\begin{tabular}{
    S[table-format=3.0]
    S[table-format=4.0] @{\hskip \TabColSep}
    S[table-format=5.0]
    S[table-format=5.0]
    S[table-format=4.0] @{\hskip \TabColSep}
    S[table-format=5.0]
    S[table-format=5.0]
    S[table-format=4.0] @{\hskip \TabColSep}
    S[table-format=2.1]}
    \toprule
    \multicolumn{2}{c @{\hskip \TabColSep}}{Instance} & 
    \multicolumn{3}{c @{\hskip \TabColSep}}{\AbbrIlpRef{}~\cite{2017:Aghelinejad}} &
    \multicolumn{3}{c @{\hskip \TabColSep}}{\AbbrIlpOur{}} &
    \multicolumn{1}{c}{\AbbrPreProc{}}  \\
    \cmidrule(lr{\TabColSep}){1-2} \cmidrule(lr{\TabColSep}){3-5} \cmidrule(lr{\TabColSep}){6-8} \cmidrule{9-9}
    \multicolumn{1}{c}{\scriptsize $\NumJobs$} &
    \multicolumn{1}{c @{\hskip \TabColSep}}{\scriptsize $\NumIntervals$} &
    \multicolumn{1}{c}{\scriptsize \AbbrObjective{} [-]} & \multicolumn{1}{c}{\scriptsize \AbbrLb{} [-]} & \multicolumn{1}{c @{\hskip \TabColSep}}{\scriptsize \AbbrTime{} [\si{\second}]} &
    \multicolumn{1}{c}{\scriptsize \AbbrObjective{} [-]} & \multicolumn{1}{c}{\scriptsize \AbbrLb{} [-]} & \multicolumn{1}{c @{\hskip \TabColSep}}{\scriptsize \AbbrTime{} [\si{\second}]} &
    \multicolumn{1}{c}{\scriptsize \AbbrTime{} [\si{\second}]} \\
    \midrule
    150 & 527 & \Optimum{8582} & 8567 & \AbbrTimeLimit{} & \Optimum{8582} & \Optimum{8582} & 187 & 1.0 \\
    150 & 647 & 8726 & 8240 & \AbbrTimeLimit{} & \Optimum{8409} & \Optimum{8409} & 277 & 2.9 \\
    150 & 767 & 8557 & 7787 & \AbbrTimeLimit{} & \Optimum{8132} & \Optimum{8132} & 624 & 5.5 \\
    150 & 888 & 8976 & 6780 & \AbbrTimeLimit{} & \Optimum{8078} & \Optimum{8078} & 511 & 9.1 \\
    170 & 650 & 10596 & 9628 & \AbbrTimeLimit{} & \Optimum{10068} & \Optimum{10068} & 290 & 2.3 \\
    170 & 799 & 10794 & 8832 & \AbbrTimeLimit{} & \Optimum{9820} & \Optimum{9820} & 1087 & 4.6 \\
    170 & 948 & 10940 & 8343 & \AbbrTimeLimit{} & \Optimum{9637} & \Optimum{9637} & 806 & 9.3 \\
    170 & 1097 & 11189 & 8124 & \AbbrTimeLimit{} & \Optimum{9620} & \Optimum{9620} & 1345 & 13.4 \\
    190 & 757 & 12555 & 11206 & \AbbrTimeLimit{} & \Optimum{12008} & \Optimum{12008} & 246 & 3.9 \\
    190 & 930 & 12882 & 10521 & \AbbrTimeLimit{} & \Optimum{11758} & \Optimum{11758} & 942 & 6.9 \\
    190 & 1104 & 12791 & 9949 & \AbbrTimeLimit{} & \Optimum{11611} & \Optimum{11611} & 3147 & 13.3 \\
    190 & 1277 & 12757 & 0 & \AbbrTimeLimit{} & \Optimum{11465} & \Optimum{11465} & 1348 & 22.7 \\
    \midrule
    \multicolumn{4}{l}{Average time [\si{\second}]:} & {\textgreater 3600} & & & {901} & {7.9} \\
    \multicolumn{4}{l}{Average optimality gap [\si{\percent}]:} & {8.39} & & & {0.00} \\
    \bottomrule
    \end{tabular}
    
\end{adjustbox}

%% file: tables/experiment-large-multioff.tex
\begin{adjustbox}{max width=\textwidth}

\renewrobustcmd{\bfseries}{\fontseries{b}\selectfont}
    
\begin{tabular}{
    S[table-format=3.0]
    S[table-format=4.0] @{\hskip \TabColSep}
    S[table-format=5.0]
    S[table-format=5.0]
    S[table-format=5.0] @{\hskip \TabColSep}
    S[table-format=5.0]
    S[table-format=5.0]
    S[table-format=4.0] @{\hskip \TabColSep}
    S[table-format=2.1]}
    \toprule
    \multicolumn{2}{c @{\hskip \TabColSep}}{Instance} & 
    \multicolumn{3}{c @{\hskip \TabColSep}}{\AbbrIlpRef{}~\cite{2017:Aghelinejad}} &
    \multicolumn{3}{c @{\hskip \TabColSep}}{\AbbrIlpOur{}} &
    \multicolumn{1}{c}{\AbbrPreProc{}}  \\
    \cmidrule(lr{\TabColSep}){1-2} \cmidrule(lr{\TabColSep}){3-5} \cmidrule(lr{\TabColSep}){6-8} \cmidrule{9-9}
    \multicolumn{1}{c}{\scriptsize $\NumJobs$} &
    \multicolumn{1}{c @{\hskip \TabColSep}}{\scriptsize $\NumIntervals$} &
    \multicolumn{1}{c}{\scriptsize \AbbrObjective{} [-]} & \multicolumn{1}{c}{\scriptsize \AbbrLb{} [-]} & \multicolumn{1}{c @{\hskip \TabColSep}}{\scriptsize \AbbrTime{} [\si{\second}]} &
    \multicolumn{1}{c}{\scriptsize \AbbrObjective{} [-]} & \multicolumn{1}{c}{\scriptsize \AbbrLb{} [-]} & \multicolumn{1}{c @{\hskip \TabColSep}}{\scriptsize \AbbrTime{} [\si{\second}]} &
    \multicolumn{1}{c}{\scriptsize \AbbrTime{} [\si{\second}]} \\
    \midrule
        150 & 529 & \Optimum{21910} & 21562 & \AbbrTimeLimit{} & \Optimum{21910} & \Optimum{21910} & 130 & 1.1 \\
    150 & 649 & 29425 & 20685 & \AbbrTimeLimit{} & \Optimum{21821} & \Optimum{21821} & 702 & 3.1 \\
    150 & 769 & 37764 & 18140 & \AbbrTimeLimit{} & \Optimum{21353} & \Optimum{21353} & 949 & 5.2 \\
    150 & 890 & 43929 & 16799 & \AbbrTimeLimit{} & \Optimum{21266} & \Optimum{21266} & 701 & 8.5 \\
    170 & 651 & 28425 & 24983 & \AbbrTimeLimit{} & \Optimum{25807} & \Optimum{25807} & 809 & 2.6 \\
    170 & 799 & 39095 & 21981 & \AbbrTimeLimit{} & \Optimum{25518} & \Optimum{25518} & 1244 & 5.0 \\
    170 & 948 & 46083 & 20709 & \AbbrTimeLimit{} & \Optimum{25279} & \Optimum{25279} & 2922 & 8.5 \\
    170 & 1096 & 53177 & 20091 & \AbbrTimeLimit{} & \Optimum{25279} & \Optimum{25279} & 2162 & 14.1 \\
    190 & 756 & 38471 & 27984 & \AbbrTimeLimit{} & \Optimum{30563} & \Optimum{30563} & 797 & 4.2 \\
    190 & 929 & 46319 & 26166 & \AbbrTimeLimit{} & \Optimum{30224} & \Optimum{30224} & 1069 & 7.5 \\
    190 & 1102 & 53751 & 24630 & \AbbrTimeLimit{} & \Optimum{30224} & \Optimum{30224} & 2069 & 13.6 \\
    190 & 1275 & 61547 & 0 & \AbbrTimeLimit{} & \Optimum{30071} & \Optimum{30071} & 2572 & 23.7 \\
    \midrule
    \multicolumn{4}{l}{Average time [\si{\second}]:} & {\textgreater 3600} & & & {1344} & {8.1} \\
    \multicolumn{4}{l}{Average optimality gap [\si{\percent}]:} & {61.33} & & & {0.00} \\
    \bottomrule
    \end{tabular}
\end{adjustbox}